\documentclass[sigconf]{acmart}

\def\BibTeX{{\rm B\kern-.05em{\sc i\kern-.025em b}\kern-.08emT\kern-.1667em\lower.7ex\hbox{E}\kern-.125emX}}

\usepackage{booktabs} 
\usepackage{balance}
\usepackage{algorithm}
\usepackage{algpseudocode}

\usepackage{soul} 

\copyrightyear{2019}
\acmYear{2019}
\setcopyright{acmlicensed}
\acmConference[UMAP '19]{27th Conference on User Modeling, Adaptation and Personalization}{June 9--12, 2019}{Larnaca, Cyprus}
\acmPrice{15.00}
\acmDOI{10.1145/3320435.3320458}
\acmISBN{978-1-4503-6021-0/19/06}

\begin{document}
\title{Extending a Tag-based Collaborative Recommender with Co-occurring Information Interests}

\author{Noemi Mauro}
 \orcid{}
 \affiliation{%
   \institution{Computer Science Department, 
   University of Torino}
   \streetaddress{Corso Svizzera 185}
   \city{Torino} 
   \state{Italy} 
   \postcode{10149}
 }
\email{noemi.mauro@unito.it}

\author{Liliana Ardissono}
 \orcid{}
 \affiliation{%
   \institution{Computer Science Department, 
   University of Torino}
   \streetaddress{Corso Svizzera 185}
   \city{Torino} 
   \state{Italy} 
   \postcode{10149}
 }
\email{liliana.ardissono@unito.it}

\renewcommand{\shortauthors}{}

\begin{abstract}
Collaborative Filtering is largely applied to personalize item recommendation but its performance is affected by the sparsity of rating data. In order to address this issue, recent systems have been developed to improve recommendation by extracting latent factors from the rating matrices, or by exploiting trust relations established among users in social networks. 

In this work, we are interested in evaluating whether other sources of preference information than ratings and social ties can be used to improve recommendation performance. Specifically, we aim at testing whether the integration of frequently co-occurring interests in information search logs can improve recommendation performance in User-to-User Collaborative Filtering (U2UCF).
For this purpose, we propose the \textit{Extended Category-based Collaborative Filtering (ECCF)} recommender, which enriches category-based user profiles derived from the analysis of rating behavior with data categories that are frequently searched together by people in search sessions. 
We test our model using a big rating dataset and a log of a largely used search engine to extract the co-occurrence of interests. 
The experiments show that ECCF outperforms U2UCF and category-based collaborative recommendation in accuracy, MRR, diversity of recommendations and user coverage. Moreover, it outperforms the SVD++ Matrix Factorization algorithm in accuracy and diversity of recommendation lists.
\end{abstract}



\maketitle

\section{Introduction}
Recommender systems research has employed item ratings, bookmarking actions and other user activities as primary sources of information to generate personalized suggestions because they provide evidence about user preferences. 
In particular, User-to-User Collaborative Filtering \cite{Desrosiers-Karypis:11} (henceforth, denoted as U2UCF) analyzes the ratings of items provided by users in order to identify ``like-minded'' people for preference prediction. However, the sparsity of the rating matrices affects recommendation performance. Thus, recent algorithms have been proposed to improve the recognition of preference similarity from rating data (e.g., Matrix Factorization algorithms \cite{,Koren-Bell:11} such as SVD++ \cite{Koren:08}), possibly combined with trust information derived from the establishment of social links among users; e.g., \cite{Tang-etal:13,Yang-etal:17}. While these algorithms achieve good accuracy and coverage, they challenge the explanation of recommendation results because the policies applied to rank items can hardly be described in an intuitive way.

In the present work, we are interested in assessing whether U2UCF, which has nice explanation properties, can be improved by using other types of information that are complementary to rating data.
Specifically, we investigate whether the identification of frequently co-occurring interests in information search can be used to improve recommendation performance. 
We start from the observation that, if the people who search for items tagged with a certain information category typically also search for items tagged with another category, the two categories might represent related interests.
Therefore, even though we ignore the reasons behind this relatedness, we might leverage the strength of the association in preference estimation.
In this perspective, we propose to to build rich user profiles by extending the preferences for categories of items identified from rating behavior with frequently co-occurring interests for item categories, extracted from the logs of search engines. It can be noticed that interest co-occurrence can be learned by analyzing anonymous interaction sessions because it is aimed at describing general user behavior. Therefore, it can be applied to anonymized search logs, as long as search sessions can be identified.

Starting from a category-based representation of user preferences, based on the analysis of ratings and on items categorization, we propose the following research question:  

{\em RQ: How does the integration of data about interest co-occurrence in information search influence the performance of a collaborative recommender system that manages category-based user profiles? }

In order to answer this question we start from a {\em Simple Category-based Collaborative Filtering (SCCF)} algorithm which infers a user's preferences on the basis of the distribution of her/his ratings on item categories: a category-based user profile provides a conceptual view on preferences, so that user similarity can be computed by abstracting from item ratings, thus contrasting data sparsity; see \cite{Sieg-etal:07b,Sieg-etal:10b}.
Then, we propose the {\em Extended Category-based Collaborative Filtering (ECCF)} algorithm that enriches category-based user profiles with evidence about interests that frequently co-occur in information search. ECCF employs the extended user profiles for rating estimation.

In order to evaluate the recommendation performance of ECCF, we extract information about co-occurring interests by analyzing the query log of a largely used search engine. Then, we test our algorithm by applying it to the Yelp Dataset \cite{Yelp-dataset}, which stores user ratings of various types of businesses.

We analyze a few settings of ECCF in order to integrate different amounts of information about co-occurring preferences with rating data. In our experiments, we evaluate performance by taking U2UCF and SCCF as baselines: these algorithms differ in neighbor identification but are based on the same rating estimation approach. Therefore, they are a good basis to assess the impact of extended category-based user profiles on preference prediction. We also compare these algorithms with SVD++ to evaluate whether preference extension challenges the capability of recommending relevant items.
The results of our experiments show that ECCS outperforms U2UCF and SCCF in accuracy, MRR, diversity of recommendations and user coverage; moreover it outperforms SVD++ in accuracy and diversity of the generated suggestion lists. We thus conclude that preference co-occurrence information can positively contribute to the identification of good neighbors for rating estimation.

In summary, the main contributions of this work are:
\begin{itemize}
    \item
    The integration of data about frequently co-occurring information interests (inferred by observing general search behavior) with category-based user preferences, in order to acquire rich individual user profiles.
    \item The ECCF category-based recommendation algorithm, which extends User-to-User Collaborative Filtering to take both frequently co-occurring information interests and preference similarity into account in neighbor identification.
    \item Evaluation results aimed at proving the benefits of frequently co-occurring interests to Collaborative Filtering.
\end{itemize}

In the following,
Section \ref{sec:related} positions our work in the related one. Section \ref{model} presents ECCF. Section \ref{sec:validation} describes the experiments we carried out to validate ECCF and discusses the evaluation results. Section \ref{sec:conclusions} concludes the paper and outlines our future work.

\section{Related Work}
\label{sec:related}

\subsection{Recommender Systems}

Cross-domain recommendation has received the researchers' attention as a way to employ multiple information sources to contrast data sparsity; e.g., \cite{Fernandez-Tobias-etal:16}. Moreover, holistic user models have been developed that jointly analyze different types of user behavior to enhance the recognition of the user's needs; e.g., \cite{Teevan-etal:05, Musto-etal:2018b}.
However, the fusion of personal information from different applications is problematic, unless it is done within a tightly integrated software environment. For instance, most people operate anonymously \cite{Greenstein-etal:17} or have multiple identities \cite{Doychev-etal:14}; moreover, most user activity logs are anonymized for privacy preservation purposes. It is thus interesting to consider other types of knowledge integration that do not require user identification across applications. Our work investigates this path of research.

Collaborative Filtering generates suggestions by analyzing item ratings to identify similar users or similar items.
Several algorithms have been developed, from K-Nearest Neighbors (KNN) to more recent ones such as Matrix Factorization \cite{Desrosiers-Karypis:11,Koren-Bell:11}. In our work we adopt KNN because it has nice explanation capabilities and has proved to achieve good performance in a comparison with other approaches  \cite{Jannach-Ludewig:17,Ludewig-Jannach:18}. 

Ontological user profiles model preferences at the semantic level. In \cite{Sieg-etal:07b,Sieg-etal:10b}, Sieg et al. propose to exploit a taxonomy whose concepts represent item types, and to infer user interests on the basis of the observed ratings to the instances of such concepts. The neighborhood for rating estimation is then identified by measuring the semantic similarity between ontological user profiles. The category-based user similarity we propose is close to this approach. However, we go one step forward in the identification of preferences by extending the user profiles with frequently co-occurring information interests.
This type of extension also differentiates our work from that of Ronen et al., who propose to extend the preferences of the individual user by analyzing her/his behavior in search logs \cite{Ronen-etal:16}: that work assumes that the user's activities can be tracked across applications and extends the user profile by analyzing her/his overall behavior. In contrast, we extend user preferences by analyzing anonymous data about general search behavior. 

Sen et al. define tag-aware recommender systems as
``recommender algorithms that predict user's preferences for tags''.
In \cite{Sen-etal:09} they describe different signs of interest; e.g., searching or applying a tag, and so forth.
Our work relates to tag-aware recommender systems because we analyze rating behavior on items associated to categories expressed as tags. However, we do not consider any other types of interaction with tags for estimating user preferences. 

In \cite{Gemmel-etal:12}, Gemmel et al. present a linear-weighted hybrid framework for resource recommendation that models different scenarios, among which tag-specific item recommendation. They propose to match users and items on the basis of their tag profiles. Differently, we match users on the basis of category-based profiles learned from rating behavior.
The same kind of difference holds between our work and the one of Nakamoto \cite{Nakamoto:2007}.

While TagiCoFi \cite{Zhen:2009} employs user similarities defined from tagging information to regularize Matrix Factorization, we use tags in a KNN algorithm. 
In \cite{Tso-Sutter:2008} Tso and Sutter extend the ratings matrix using tagging information.
They reduce the three-dimensional correlations $<user, tag, item>$ to two-dimensional correlations $<user,tag>$, $<item,tag>$ and $<user, item>$. Then, they apply a fusion method to combine the correlations for rating prediction. Differently, we extend the rating matrix with the categories (tags) associated to the items rated by users and with further categories identified from general search behavior. 

Recently, rating information has been combined with other types of data to improve recommendation. For instance, item reviews are used, possibly in combination with ratings, in \cite{Chen-etal:15,Musat-Faltings:15,Muhammad-etal:15,Lu-etal:18}.
Moreover, trust relations and reputation are used to steer recommendation on the basis of the feedback on items provided by trusted parties; e.g., \cite{Kuter-etal:07,Liu-Lee:10,Tang-etal:13,Alotaibi-Vassileva:16,Mcnally-etal:14,Du-etal:17,Yang-etal:17}.
In \cite{Mauro-etal:19}, we investigate multi-faceted trust for personalized recommendation. 
However, in the present work we focus on rating information to assess the potential improvement of Collaborative Filtering, when combined with general preference co-occurrence.

\subsection{Analysis of Interaction Sessions}
The identification of interest co-occurrence we propose is related to a few works supporting query expansion, query reformulation and term suggestion in Information Retrieval. Some researchers propose to analyze session-based user behavior in order to detect co-occurrence relations useful to improve search queries, taking the search context into account. For instance, in \cite{Cao-etal:08} Cao et al. suggest queries on the basis of the context provided by the user's recent search history, by clustering queries on the basis of the search results visited by users. Moreover, Huang et al. \cite{Huang-etal:03} and Chen et al. \cite{Chen-etal:08} detect term co-occurrence in search sessions to group sets of relevant words that can be mutually suggested. 
Our work is different because we adopt a linguistic interpretation approach (based on lemmatization and Word Sense Disambiguation) to find the concepts referenced in the queries; see \cite{Mauro-Ardissono:17b}. 
Therefore, we extract information about {\em concept co-occurrence}, which is more general than {\em term co-occurrence}. 

It is worth mentioning that our analysis of interaction sessions differs from session-based recommendation, which analyzes the user's behavior during an interaction session to identify relevant item(s) to suggest; e.g., see \cite{Garcin-etal:13,Jannach-Ludewig:17,Greenstein-etal:17,Jannach-etal:17}. In fact, we mine interest co-occurrence by abstracting from the particular sequence of queries performed by the users. Moreover, as previously discussed, we mine concept associations.

\subsection{Graph-based Information Filtering}
Knowledge graphs describe item features and relations among entities, supporting the analysis of item relatedness, as well as similarity for information filtering and top-N recommendation. 
In several works these graphs are extracted from document pools and/or from the Linked Data Cloud. For instance, CoSeNa \cite{Candan-etal:09} employs keyword co-occurrence in the corpus of documents to be retrieved, and ontological knowledge about the domain concepts, to support the exploration of text collections using a keywords-by-concepts graph. Moreover, in \cite{DiNoia-etal:16}, Di Noia et al. create a relatedness graph by analyzing external data sources such as DBpedia in order to support the evaluation of semantic similarity between items. Analogously, item features have been extracted from the Linked Data Cloud to improve recommendation performance in \cite{Musto-etal:16,Ragone-etal:17,Musto-etal:17,Musto-etal:18}. 

Some works attempt to extend the relations among information items by integrating data derived from the observation of different types of user behavior. E.g., Google search engine manages the Knowledge Graph \cite{GoogleKnowledgeGraph} to relate facts, concepts and entities depending on their co-occurrence in queries. Moreover, entity2rec learns user-item relatedness from knowledge graphs by analyzing data about users' feedback and item information from Linked Open Data \cite{Palumbo-etal:17}. Furthermore, in \cite{Oramas-etal:15} Oramas et al. propose a hybrid recommender that integrates users implicit feedback into a knowledge graph describing item information, enriched with semantic data extracted from external sources. Finally, in \cite{Vahedian-etal:17}, Vahedian et al. generalize graph-based approaches by simultaneously taking into account multiple types of relations among entities: they introduce meta-paths to represent patterns of relations and apply random-walk along such paths to identify relevant entities to suggest.

Our work has analogies to the above listed ones because we employ a graph-based type of knowledge representation. However, we work at the conceptual level: our knowledge graph relates item categories instead of individual users and/or items. Moreover, we do not compute similarity or relatedness by means of the knowledge graph: we use the graph to extend category-based user profiles. In turn, those profiles are employed in neighborhood identification. The separation between how preferences are inferred and how they are used for recommendation makes it possible to extend both types of activities in a modular way.

\section{Extended Category-based Collaborative Filtering}
\label{model}
We describe ECCF incrementally, starting from U2UCF that provides the basic match-making approach for rating estimation. 

\subsection{User-to-User Collaborative Filtering}
In \cite{Ricci-etal:11}, Ricci et al. define U2UCF as follows: ``the simplest and original implementation of this approach recommends to the active user the items that other users with similar tastes liked in the past. The similarity in taste of two users is calculated based on the similarity in the rating history of the users".
Given: 
\begin{itemize}
    \item $U$ as the set of users and $I$ as the set of items;
    \item $r: U X I \Rightarrow {\rm I\!R}$ as a map of ratings;
    \item $R \in {\rm I\!R}^{U X I}$ as the users-items rating matrix, where each value is a rating $r_{ui}=R[u,i]$ given by a user $u \in U$ to an item $i \in I$. 
\end{itemize}
The recommender system estimates $u$'s rating of $i$ ($\hat{r}_{ui}$) as follows:
        \begin{equation}
        \label{eq:rmeancentering}
        \hat{r}_{ui} = \bar{r}_u + \frac{ 
        	\sum\limits_{v\in N_i(u)}\sigma(u,v) (r_{vi} - \bar{r}_v)
        }{
        	\sum\limits_{v\in N_i(u)}|\sigma(u,v)|}
    \end{equation}
where $N_i(u)$ is the set of neighbors of $u$ that rated item $i$ and $\sigma(u,v)$ is the similarity between user $u$ and user $v$ ($v \in N_i(u)$). The similarity among users is computed by applying a distance metric, e.g., Cosine or Pearson similarity, to their rating vectors.

\subsection{Simple Category-based Collaborative Filtering (SCCF)}
\label{category-based-CF}
SCCF manages user profiles in which the user's interest in each item category is represented as a positive number; the higher is the value, the stronger is the interest.  
We define:
\begin{itemize}
    \item $U$, $I$, $r$ and $R$ as above; 
    \item $C$ as the set of item categories;
    \item $f: U X C \Rightarrow {\rm I\!N} $ as a map between users and categories;
    \item $UC\in {\rm I\!N}^{U X C}$ as the Users-Categories matrix. For each $u \in U$ and $c \in C$, $UC[u,c]$ represents the interest of $u$ in $c$. We take as evidence of interest the {\em frequency of exploration} of a category, i.e., the frequency of interaction of the user with items associated with the category. 
\end{itemize}
Category exploration can be mapped to different types of user behavior; e.g., tagging items and searching for items by tag. We map exploration to rating behavior and we define $UC[u, c]$ as the number of ratings that $u$ has given to the items associated with $c$. 

SCCF computes user similarity on the basis of the estimated user preferences for item categories. Specifically, $\sigma(u, v)$ is defined as the Cosine similarity of the users vectors in the $UC$ matrix and it is used in Equation (\ref{eq:rmeancentering}) to estimate ratings. Thus, $\hat{r}_{ui}$ is computed on the basis of the ratings $r_{vi}$ provided by the users $v \in U$ whose preferences for categories are similar to those of $u$.

\subsection{Acquisition of Preferences Co-occurrence}
\label{graph}
In order to learn the strength of the associations between item categories in search behavior, we analyze their co-occurrence in the search sessions of a query log. By co-occurrence we mean the fact that two or more categories are referred by the queries belonging to the same session. 
In the following we summarize the analysis of category co-occurrence; see \cite{Mauro-Ardissono:18} for details.

The Category Co-occurrence Graph ($CCG$) represents category co-occurrence:
in the $CCG$, nodes represent the data categories referenced in the analyzed queries and the weight of edges represents the co-occurrence frequency of the connected categories; i.e., how many times the categories have been identified within the same search sessions.

We retrieve the categories occurring in the queries by applying a Natural Language approach that identifies the referred concepts in a flexible way, by considering synonyms and by applying Word Sense Disambiguation to resolve the meaning of words; see \cite{Ardissono-etal:16,Mauro-Ardissono:17b}. For Word Sense Disambiguation we use the Babelfy tool \cite{Babelfy}.

The $CCG$ is built as follows:
given two categories $x$ and $y$, the weight of the edge that connects them is defined as:
\begin{equation}
\label{eq1}
w_{xy}=\sum_{S\in|Sessions|} Freq_{S_{xy}}
\end{equation}
where $Freq_{S_{xy}}$ represents the evidence provided by session $S$ to the co-occurrence frequency of $x$ and $y$.
Given $S=\{Q_1, \dots, Q_n\}$,
$Freq_{S_{xy}}$ is computed as the maximum evidence of co-occurrence of $x$ and $y$ in $S$: 
\begin{equation}
\label{eq2}
Freq_{S_{xy}} = Max_{k=1}^{n}(Freq_{xy_{Q_k}}, ev_{xy_{Q_{k-1}}})
\end{equation}
where $Freq_{xy_{Q_k}}$ is the co-occurrence evidence of $x$ and $y$  provided by query $Q_k$, and $ev_{xy_{Q_{k-1}}}$ is the one provided by $Q_1, \dots, Q_{k-1}$. Similar to \cite{Mauro-Ardissono:18}, we take the maximum, and not the sum of evidence because co-occurrence could derive either from query reformulation \cite{Rieh-Xie:06}, or from the repetition of queries in click-through events of the log; see Section \ref{sec:AOLlog} that describes the query log we used.

A query $Q$ contributes to the estimation of co-occurrence as follows:
\begin{itemize}
\item 
If $Q$ contains $k$ terms ($k>=0$), each one identifying a non-ambiguous category: 
$T_1 \Rightarrow c_1, \quad \dots, \quad T_k \Rightarrow c_k$, then, for each category $c$ of $Q$:
\begin{itemize}
    \item The co-occurrence evidence between $c$ and every other category $d$ of $Q$ is $Freq_{cd_{Q}} = 1$. 
    \item The co-occurrence evidence between $c$ and every other category $e$ identified in a non-ambiguous way in the other queries of $S$ is $Freq_{ce_{Q}} = 1$. 
    \item The co-occurrence evidence between any other categories $w$ and $z$ identified in $S$ is $Freq_{wz_{Q}} = 0$.
\end{itemize}
\item 
If $Q$ contains an ambiguous term $t$ that refers to $m$ categories, the particular category the user is focusing on cannot be identified. Therefore, the co-occurrence evidence brought by $t$ is computed as above, but the assigned evidence is $\frac{1}{m}$ in order to consider the possible interpretations of $Q$, and divide evidence among ambiguous categories.
\end{itemize}

\begin{figure}[t]
    \centering
    \includegraphics[width=0.9\linewidth]{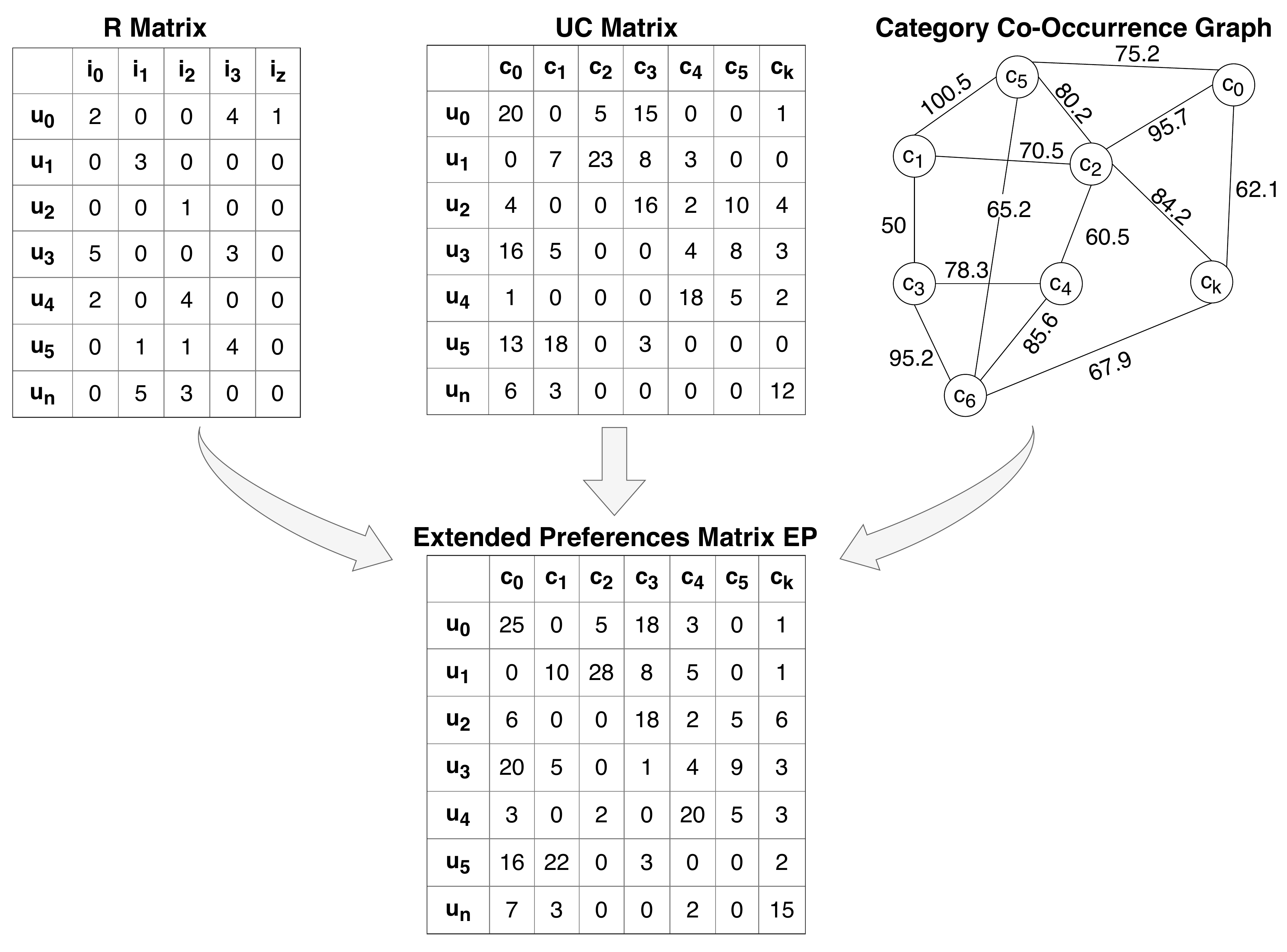}
    \caption{Extension of Category-based User Profiles. }
    \label{fig:matrici}
\end{figure}

\subsection{Extended Category-based Collaborative Filtering (\textit{ECCF})}
\label{sec:ECCF}
In this recommendation model we employ frequent co-occurring information interests to extend category-based user profiles.
We reinforce the preferences for item categories learned by analyzing rating behavior (stored in the Users-Categories matrix $UC$) with interest co-occurrence associations (stored in the $CCG$ graph) in order to acquire an extended set of user preferences for neighbor identification.

The idea behind preference extension is that, the more the user has appreciated the items of a category, the more interest
\linebreak
co-occurrence makes sense. Therefore, starting from the category-based user profiles stored in the $UC$ matrix, we increment user preferences with the contribution of the strongest co-occurrence relations of the $CCG$ graph, depending on the number of positive ratings available in the users-items matrix $R$. 
The output of this process is stored in the Extended Preferences matrix $EP$, which is used to compute $\sigma(u,v)$ in Equation \ref{eq:rmeancentering}.

Figure \ref{fig:matrici} provides a graphical view of the computation of $EP$: the information stored in $UC$ is combined with that stored in the $CCG$ to set the values of this matrix. In this process, the users-ratings matrix $R$ is used to limit the reinforcement of preferences to the categories of the positively rated items.
Moreover, the $CCG$ is used to propagate preference information according to the strongest co-occurrence of interests. 
In detail, we compute the values of $EP$ as follows:
\begin{itemize}
    \item 
    let $Cat_i$ be the set of categories associated to item $i$; 
    \item
    let $CatSet_i$ be the set of categories directly connected to any category $c \in Cat_i$ in the $CCG$ through the heaviest outbound arcs. These are the categories which most frequently co-occur with some categories of $Cat_i$ in search sessions.
\end{itemize}
Then:
\begin{equation}
\label{eq:pm}
EP[u,c]=UC[u,c]+\sum_{i\in|I|} f(u,i,c)
\end{equation}
where
\begin{equation}
             f(u,i,c) =
            \begin{cases}
                1 & \quad \text{if $R[u,i] \in$} ~ \text{\textit{PositiveRatings}} ~ \text{$\wedge$} ~\text{$c \in CatSet_i$}\\
                0 & \quad \text{otherwise}
            \end{cases}
\label{eq:f}
\end{equation}
In Equation \ref{eq:f} $PositiveRatings$ denotes the set of ratings that are considered as positive in the dataset; e.g., \{5\}, or \{4, 5\} in a [1, 5] Likert scale.

\section{Validation of ECCF}
\label{sec:validation}

\subsection{Dataset of Item Ratings}
\label{sec:YELP}
As a source of rating data we exploit the Yelp Dataset \cite{Yelp-dataset}, which contains information about a set of businesses, users and reviews and is available for academic purposes. In the dataset, item ratings take values in a [1, 5] Likert scale where 1 is the worst value and 5 is the best one. 
Moreover, each item is associated with a list of categories describing the kind of service it offers.

The full list of Yelp categories is available at \url{www.yelp.com/developers/documentation/v3/category_list} and is organized in a taxonomy to specify businesses at different levels of detail. The taxonomy includes a large set of first-level categories, representing broad types of businesses; e.g., ``Active life'', ``Arts \& entertainment'', ``Automotive'', \dots, ``Food'', ``Restaurants'', and many others. In turn, the first-level categories are specialized into sub-categories; e.g., ``Restaurants'' includes many types of restaurants such as ``Indian'', ``Chinese'' and the like.  
We apply two filters to the dataset:
\begin{enumerate}
    \item 
    We select all the Yelp categories that are subclasses of ``Restaurants'' or ``Food'': e.g., ``Indian'', ``Chinese'', ``Cafes'', ``Kebab'', ``Pizza'', and so forth; the total number of categories is 254. 
    Then, we project the Yelp dataset on the set of items associated with at least one of these categories. In the rest of this paper we refer to this set of categories as {\em CATS}. 
    \item We further filter the dataset on the users who rated at least 20 items.
\end{enumerate}

\begin{table}[t]
\centering
\caption{Statistics about the Filtered Datasets}
\begin{tabular}{l|l|l}
\hline
Yelp & Number of users            & 26,600   \\
& Number of businesses       & 76,317   \\ 
& Number of ratings          & 1,326,409 \\ \hline
AOL & Number of sessions & 1,248,803 \\
& Number of queries & 2,136,029 \\
\hline
\end{tabular}
\label{t:dataset}
\end{table}
The higher portion of Table \ref{t:dataset} summarizes the number of users, businesses and ratings of the filtered Yelp dataset.

\subsection{Dataset of Search Sessions}
\label{sec:AOLlog}
For the generation of the Category Co-occurrence Graph we use the AOL query log.\footnote{The log is available at \url{http://www.cim.mcgill.ca/~dudek/206/Logs/AOL-user-ct-collection/}.}  
Each line of the log represents either a query or a click-through event on one of the search results of a query. The line contains various fields, among which the submitted query and the submission date and hour.

In order to build a graph that is thematically related to the items of the filtered Yelp dataset, we select from the log the search sessions relevant to the categories $c \in CATS$ enriched with the following two types of external knowledge. The enrichment is useful to abstract from the specific category names used in Yelp and to take into account semantically related information: 
\begin{enumerate}
    \item 
    {\em Lemmatized knowledge:} we enrich each element $c \in CATS$ with a set of keywords and synonyms from WordNet \cite{WordNet} lexical database. 
    \item 
    {\em Relevant terms from the Probase \cite{Wu-etal:12} taxonomy:} 
    \begin{itemize}
        \item
        For each element $c \in CATS$, we enrich $c$ with the 
        \linebreak
        $<concept, instance>$ pairs of ProBase such that $concept$ has at least 85\% WordNet similarity with any term of the lemmatized knowledge of $c$, and the WordNet similarity between the two components of the pair is 85\%.
        \item 
        ProBase, recently called Microsoft Concept Graph, is a large concept network harnessed from web pages and search logs. It is organized as a list of $<instance, concept>$ pairs related by a subclass relation and it contains 
        \linebreak 
        5,376,526 classes and 12,501,527 instances.
    \end{itemize}
\end{enumerate}
For the selection of relevant search queries in the AOL log we match the lemmatized words occurring in the queries to the enriched categories of $CATS$. If there is at least one match between a term and a query, we consider the query as relevant and we include its parent session in the filtered log. 

We identify the search sessions by aggregating the queries performed by the same user according to their temporal proximity, following the widely applied rule that two consecutive queries belong to different sessions if the time interval between them exceeds half an hour; see \cite{White-etal:07}. 
 
The lower portion of Table \ref{t:dataset} shows the number of sessions and queries of the filtered AOL dataset.

It is worth noting that the AOL log was involved in an information leak issue but we decided to use it for two reasons. Firstly, our analysis is ethically correct because we study general search behavior to acquire aggregate data abstracting from the search histories of individual users. Secondly, to the best of our knowledge, the AOL log is the only available large dataset that reports textual search queries, and which can therefore be used for linguistic interpretation. We analyzed some public datasets but they did not meet our requirements. For instance, the Excite query dataset\footnote{\url{https://svn.apache.org/repos/asf/pig/trunk/tutorial/data/}} contains about 1M queries while AOL log contains 20M queries. Moreover, in the Yahoo dataset\footnote{\url{https://webscope.sandbox.yahoo.com/catalog.php?datatype=l\&did=50}} the queries are coded; thus, it is not possible to extract any linguistic information to learn category co-occurrence.

\begin{figure}
  \centering
  \includegraphics[width=0.6\linewidth]{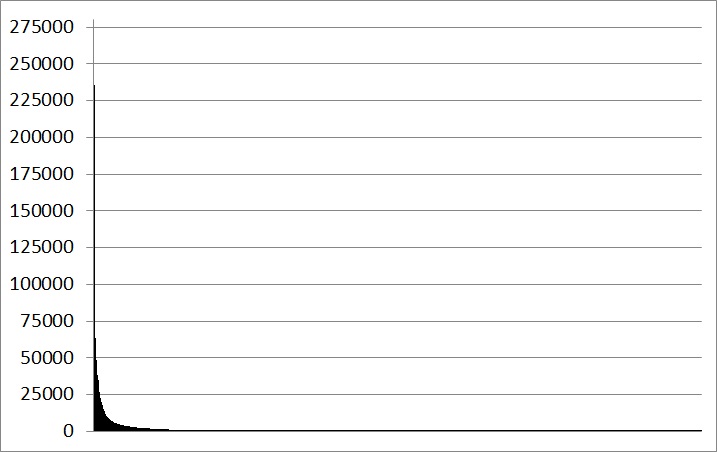}
  \caption{Distribution of the Weight of Edges in the $CCG$.}\label{fig:graph-distribution}
\end{figure}

\subsection{Category Co-occurrence Graph}
We instantiate the {\em CCG} with the interests that co-occur in the sessions of the filtered AOL dataset by applying the procedure described in Section \ref{graph}.
The resulting graph is strongly connected: almost all of the categories are linked to each other by an edge having weight $>0$.
However, the distribution of weights in the graph shows that there is a large number of weakly connected categories and a very small number of strongly associated ones. The ``heavy'' edges identify the interests that co-occur very frequently in search sessions and suggest to select the arcs having maximum weight in the {\em CCG} for the extension of the user profiles, as done in Section \ref{sec:ECCF}.
Figure \ref{fig:graph-distribution} shows this distribution; the x-axis represents the edges of the graph, and the y-axis represents their weights, which take values in [1, 272224].

\begin{table}[b]
\centering
\caption{Performance Evaluation @10; the Best Values Are in Boldface, the Worst Ones Are Strikethrough}
\begin{tabular}{l|l|l|l|l|l}
\hline 
\textbf{Metrics} 
                             & \multicolumn{1}{c|}{\textbf{\begin{tabular}[c]{@{}c@{}}U2UCF\end{tabular}}} & \multicolumn{1}{c|}{\textbf{SCCF}} & \multicolumn{1}{c|}{\textbf{\begin{tabular}[c]{@{}c@{}}ECCF\\ \{3,4,5\}\end{tabular}}} & \multicolumn{1}{c|}{\textbf{\begin{tabular}[c]{@{}c@{}}ECCF\\ \{4,5\}\end{tabular}}} & \multicolumn{1}{c}{\textbf{\begin{tabular}[c]{@{}c@{}}ECCF\\ \{5\}\end{tabular}}} \\ \hline
\textbf{Precision}    & \st{0.7823}   & \textbf{0.786}  & 0.7857   & 0.7855  & 0.7859   \\
\textbf{Recall}       & \st{0.7473}   & 0.7526   & 0.7536   & \textbf{0.755}   & 0.7529  \\
\textbf{F1}           & \st{0.7644}   & 0.7689   & 0.7693   & \textbf{0.7699}  & 0.769   \\
\textbf{RMSE}         & \st{1.0001}   & 0.9899   & 0.9897   & 0.9893 & \textbf{0.9892}   \\
\textbf{MRR}          & \st{0.733}    & 0.7367   & 0.737    & \textbf{0.7391}  & 0.7384    \\ 
\textbf{Diversity}  & \st{0.3042}  & 0.3053 & \textbf{0.3056} & 0.3053  & 0.3049 \\
\textbf{User cov.}   & \st{0.8497}  & 0.8521  & 0.8526 & \textbf{0.8542}  & 0.8534  \\
\hline
\end{tabular}%
\label{t:results@10}
\end{table}

\subsection{Test Methodology}
\label{experiments}
We evaluate the recommendation performance of ECCF by comparing it to U2UCF and SCCF, which we consider as baselines. Moreover, we compare these algorithms with SVD++ in order to assess the improvement in the suggestion of relevant items given by frequently co-occurring interests.

The SCCF and ECCF recommendation algorithms are developed by extending the Surprise library \cite{Surprise}, while we use the default Surprise implementations of U2UCF and SVD++.

We test the algorithms by applying a 10-fold cross-validation on the filtered Yelp dataset, after having randomly distributed ratings on folds: we use 90\% of the ratings as training set and 10\% as test set. In all the tests, we configure the KNN algorithms to work with 50 neighbors.

In order to analyze the impact on recommendation performance of a looser, or stricter extension of user preferences with category co-occurrence, we validate ECCF on different settings of $PositiveRatings$ in Equation \ref{eq:f}, i.e., on different interpretations of what is a good rating. For each fold
we generate three versions of the Extended Preferences matrix $EP$ having set $PositiveRatings$ to $\{3,4,5\}$, $\{4,5\}$, and $\{5\}$ respectively.  

We evaluate Top-k recommendation performance with k=10 and k=20 by taking the ratings observed in the Yelp dataset as ground truth. For the evaluation we consider the following metrics: Precision, Recall, F1, RMSE, MRR, Diversity and User Coverage.

Diversity describes the mean intra-list diversity of items in the suggestion lists @k; see \cite{Bradley-Smyth:01}. In this work, we interpret diversity from the viewpoint of item classification. Therefore, we measure the diversity of a recommendation list as follows:
\begin{equation}
\text{intra-list diversity@k}={\frac  {\sum _{{i=1}}^{k}\sum _{{j=i}}^{k} (1 - sim(i, j))} {\frac{k*(k+1)}{2}}}
\end{equation}
where $sim(i, j)$ is the cosine similarity between the lists of categories associated to items $i$ and $j$ in the ratings dataset. 

\subsection{Results}
\label{results}
Table \ref{t:results@10} shows the performance results of the KNN recommenders we compared, by taking into account a maximum of 10 suggested items (performance@10). 
\begin{itemize}
    \item \textbf{Precision:} similar to previous results described in \cite{Sieg-etal:07b}, all of the category-based recommenders outperform U2UCF. This can be explained by the fact that the matrices describing preferences for item categories are denser than the ratings one. Thus, they improve recommendation by supporting a better identification of neighbors for Equation \ref{eq:rmeancentering}.  
    However, SCCF outperforms all of the ECCF variants. The second best recommender is ECCF$\{5\}$ that extends user profiles in the strictest way: it only considers as pivots for extension the categories associated to the items that the user has rated 5 stars. Notice also that the precision of ECCF decreases when $PositiveRatings$ is lax. The reason is that the extension of user profiles with frequently co-occurring interests can increase the estimated interest in some noisy categories with respect to the pure observation of ratings distribution on categories. In particular, noise grows when the policy applied to extend preferences is less restrictive. 
    \item \textbf{Recall:} ECCF outperforms the baselines in all the settings of $PositiveRatings$. Specifically, ECCF\{4,5\} achieves the best result, while recall is lower in ECCF\{3,4,5\} and further decreases in ECCF\{5\}.
    We explain this finding as follows: an extension of user profiles based on the categories of highly rated items supports the identification of a richer set of user preferences, and a more efficacious identification of neighbors, than only considering rating distribution on categories. However, if we restrict $PositiveRatings$ too much, the user profiles are not extended enough to sensibly improve Recall. Moreover, as noticed for Precision, if $PositiveRatings$ is lax, noise in the estimation of user preferences challenges neighbor selection.
    
\begin{table}[t]
\centering
\caption{Performance Evaluation @20}
\begin{tabular}{l|l|l|l|l|l}
\hline 
\textbf{Metrics} 
                             & \multicolumn{1}{c|}{\textbf{\begin{tabular}[c]{@{}c@{}}U2UCF\end{tabular}}} & \multicolumn{1}{c|}{\textbf{SCCF}} & \multicolumn{1}{c|}{\textbf{\begin{tabular}[c]{@{}c@{}}ECCF\\ \{3,4,5\}\end{tabular}}} & \multicolumn{1}{c|}{\textbf{\begin{tabular}[c]{@{}c@{}}ECCF\\ \{4,5\}\end{tabular}}} & \multicolumn{1}{c}{\textbf{\begin{tabular}[c]{@{}c@{}}ECCF\\ \{5\}\end{tabular}}} \\ \hline
\textbf{Precision}  & \st{0.7806} & \textbf{0.7842}  & 0.7839   & 0.7838  & \textbf{0.7842} \\ 
\textbf{Recall}     & \st{0.757}  & 0.7624  & 0.7634  & \textbf{0.7649} & 0.7626  \\ 
\textbf{F1}  & \st{0.7686}  & 0.7731   & 0.7735  & \textbf{0.7742}  & 0.7732  \\
\textbf{RMSE}  & \st{0.9935}  & 0.9838 & 0.9835 & \textbf{0.9832}  & \textbf{0.9832} \\ 
\textbf{MRR}  & \st{0.733} & 0.7369   & 0.7372   & \textbf{0.7391}  & 0.7384  \\ 
\textbf{Diversity} & \st{0.3059}  & 0.307  & \textbf{0.3073}  & 0.307  & 0.3067  \\ 
\textbf{User cov.}   & \st{0.8497}  & 0.8521  & 0.8526 & \textbf{0.8542}  & 0.8534  \\
\hline
\end{tabular}%
\label{t:results@20}
\end{table}

\begin{figure}[b]
 \centering
 \includegraphics[width=0.9\linewidth]{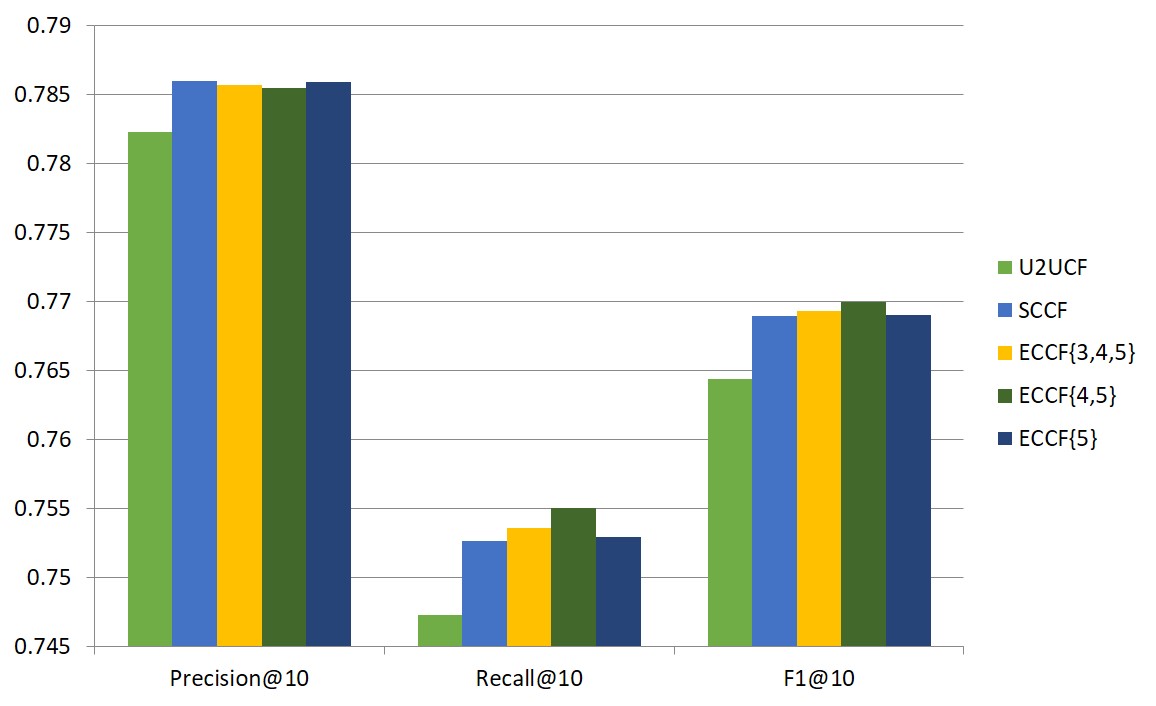}
 \caption{Graphical Representation of Accuracy@10.}
 \label{fig:accuracy@10}
\end{figure}

    \item \textbf{F1:} ECCF outperforms the baselines. In detail, ECCF\{4,5\} achieves the best F1 = 0.7691; moreover, F1 varies consistently with Recall, depending on $PositiveRatings$. 
    \item \textbf{RMSE:} SCCF reduces the mean error between estimated and observed ratings with respect to the baseline, showing the benefits of category-based user profiles. Moreover, consistently with the variation of Precision, the best results are obtained by ECCF\{5\}, i.e., with a strict extension of user profiles. RMSE progressively increases (i.e., gets worse) for $PositiveRatings=\{4, 5\}$ and \{3, 4, 5\}.
    \item \textbf{MRR:} ECCF outperforms the baselines. Specifically, 
    \linebreak
    ECCF\{4,5\} obtains the best MRR = 0.7391. The second best value corresponds to a more selective extension of user profiles in ECCF\{5\}; moreover, if $PositiveItems=\{3, 4, 5\}$ results get worse.
    \item \textbf{Diversity}: both SCCF and ECCF outperform U2UCF. In this case, the best results are obtained with a lax extension of user preferences (ECCF\{3,4,5\}) and Diversity decreases while the preference extension policy becomes stricter. We explain these findings with the fact that category-based user profiles improve the estimation of user preferences concerning a variegate set of item categories, with respect to a flat recommendation based on ratings. However, the stricter is the extension of user preferences, the less item categories are used in neighbor identification.
     \item \textbf{User coverage:} ECCF outperforms the baselines, confirming the usefulness of preference extension. However, the selection of the ratings for the extension influences coverage: ECCF\{4,5\} achieves the best results by suggesting at least one relevant item to 85.42\% of the users, against 84.97\% of U2UCF. The second best is ECCF\{5\} and ECCF\{3,4,5\} has the worst results.
    \end{itemize}
In the described experiments the $EP$ Matrix is defined by only taking into account positive ratings. In order to get a broader view on the performance of ECCF, we also consider its application to all the user ratings; i.e., we set $PositiveRatings$ to $\{1, \dots, 5\}$. With respect to the previous results, in this case the algorithm achieves similar Precision but lower Recall (0.7524), MRR (0.7369) and User coverage (0.8155).   

Table \ref{t:results@20} shows the results obtained by comparing 
\linebreak
performance@20. 
These results confirm the usefulness of category-based user profiles and of their extension with frequently
\linebreak
co-occurring information interests: 
\begin{itemize}
    \item Also in this case, ECCF\{4,5\} is the best recommendation algorithm. It outperforms the others in Recall, F1, MRR and User coverage. Moreover both ECCF\{5\} and ECCF\{4,5\} achieve the best RMSE in comparison with the other recommenders.  
    \item However, while SCCF has the best Precision@10, both SCCF and ECCF\{5\} achieve the best Precision@20.
\end{itemize}
With respect to k=10, Precision@20 is lower while Recall@20 and F1@20 take higher values; this makes sense because we are considering longer suggestion lists. Moreover, RMSE@20 is lower, which tells us that the longer lists contain proportionally less errors in the estimation of ratings. Differently, most algorithms obtain the same MRR for k=10 and k=20 (except for SCCF and ECCF\{3,4,5\}): this shows that the first relevant item is almost always placed in the first 10 positions of the suggestion lists.
Furthermore, the Diversity@20 has the highest values for all the recommenders: this might be due to the fact that the longer suggestion lists have more chances to include items belonging to different categories. Finally, User coverage@10 = User coverage@20 because we interpret coverage as the percentage of users who receive at least one suggestion.

\begin{figure}[b]
 \centering
 \includegraphics[width=0.9\linewidth]{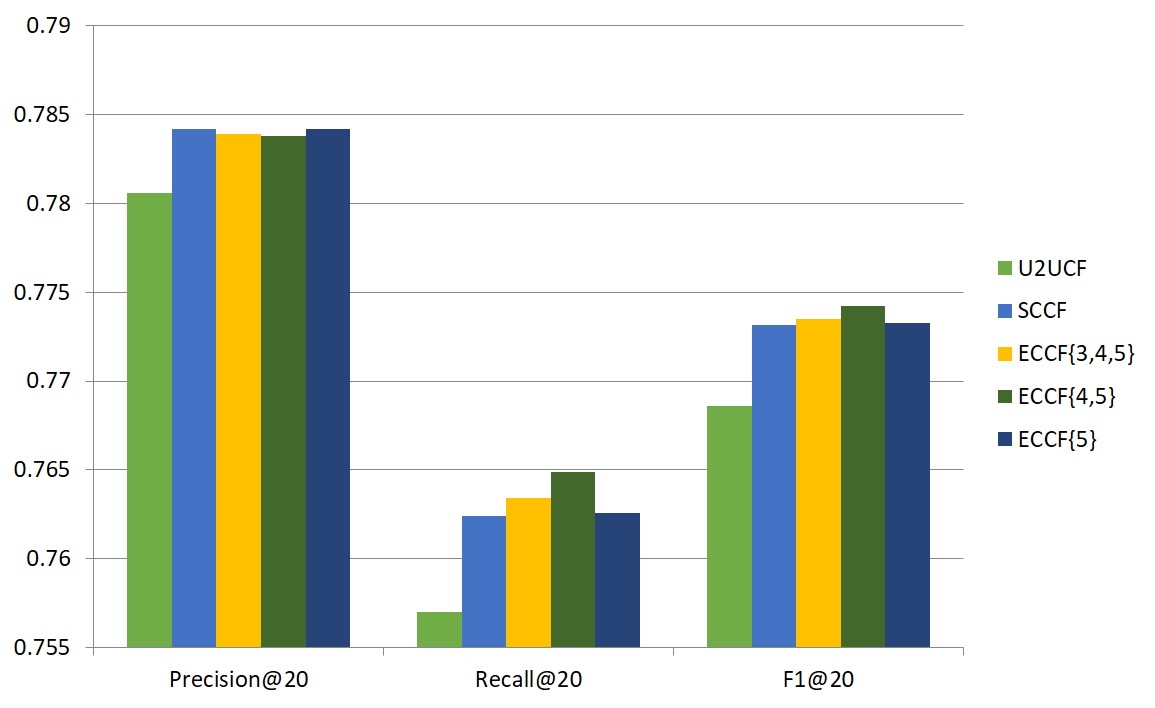}
 \caption{Graphical Representation of Accuracy@20.}
 \label{fig:accuracy@20}
\end{figure}

Figures \ref{fig:accuracy@10} and \ref{fig:accuracy@20} depict the accuracy @10 and @20:
\begin{itemize}
    \item 
    All of the category-based recommenders outperform U2UCF, confirming the benefits of the introduction of category-based preferences in KNN Collaborative Filtering. The conceptual representation of user preferences generally improves performance because the matrices describing user preferences ({\em UC} and {\em EP}) are denser than the users-items matrix storing ratings ({\em R}). Therefore, better neighbors can be identified for the computation of Equation \ref{eq:rmeancentering}.
    \item
    A comparison between category-based algorithms shows that the best performance results are obtained by extending user profiles on the basis of the items that users have rated very well, i.e., with 4 or 5 stars in a [1, 5] Likert scale. If the items that received middle ratings are considered as well, accuracy decreases. 
    \item The category-based representation of user profiles has positive impact on the Diversity of recommendation lists. Conversely, the extension of user profiles does not further help this aspect, unless user profiles are extended in a lax way. However, a lax extension is not convenient because it decreases other measures.
\end{itemize}

In order to assess the usefulness of preference extension in Top-k recommendation, we also compare the previously described algorithms with SVD++ \cite{Koren:08}, which adopts Matrix Factorization to learn latent user and item factors, basing rating prediction on the sole analysis of user ratings. The comparison results show that:
\begin{itemize}
    \item 
    SVD++ is more accurate than U2UCF and SCC. On the filtered Yelp dataset, SVD++ obtains F1@10 = 0.7696. This finding shows that the management of category-based user profiles helps recommendation but it can be outperformed by a deeper understanding of the features of items and users. 
    \item
    SVD++ achieves similar accuracy results with respect to ECCF but it is outperformed by ECCF\{4, 5\}. Therefore, the extension of user profiles with frequently co-occurring information interests, integrated into a KNN recommender, improves accuracy and makes it comparable or higher than that of Matrix Factorization algorithms. 
    \item
    ECCF outperforms SVD++ as far as the diversity of the recommendation lists is concerned: SVD++ has Diversity@10 = 0.3041; this is comparable to the diversity achieved by U2UCF and lower than that of all the category-based recommenders we presented.
    \item
    In contrast, SVD++ has the highest User coverage of all the algorithms (0.8709), showing its superior capability to contrast data sparsity. 
    \end{itemize}
    
\subsection{Discussion}
\label{discussion}
In summary, the evaluation results show that ECCF outperforms U2UCF, SCCF and SVD++ in accuracy and intra-list diversity. Moreover, it outperforms U2UCF and SCCF in MRR and user coverage, while SVD++ excels in the latter metric. The results also show that ECCF achieves the best results when applied to positive ratings, while its performance slightly decreases when the user profiles are extended by taking both positive and negative ratings.

These results support the hypothesis that preference extension, based on frequently co-occurring information interests, improves the accuracy of the suggestions generated by a KNN recommender system. However, research has to be carried out to improve other performance metrics, possibly also investigating the integration of preference co-occurrence in Matrix Factorization algorithms.

It might be questioned whether extending user profiles with general interest co-occurrence data might provide less personalized recommendations than, e.g., focusing the extensions on the user's neighborhood. In this respect, we point out that we aim at developing a model that does not depend on cross-domain user identification. However, an investigation of this issue can be interesting to deal with the cases in which user information can be shared among the applications, or public information about the users can be connected to the local profiles; e.g., public data on social networks.

Before closing this discussion, it is worth noting that, even though the AOL query log dates back to 2006, it can be considered as a good information source as long as it is analyzed from the viewpoint of the concepts expressed by the users. In other words, while the specific information items mentioned in the log might not exist any more, the topics referred in the queries are general and long-lasting. Of course, some new topics (e.g., new types of restaurants) might have emerged since 2006, and maybe new concept associations could exist now. However, the described performance results show that the co-occurring interests we identified are useful to improve recommendation performance; moreover, the methodology described in this paper can be applied to other more recent datasets, if available.

\section{Conclusions}
\label{sec:conclusions}
We investigated whether the identification of frequently
\linebreak 
co-occurring interests in information search can be used to improve the performance of KNN collaborative recommender systems. For this purpose, we defined a preference extension model that, applied to a category-based representation of user profiles, infers user preferences by exploiting frequently co-occurring information interests. Then, we implemented the model in the Extended Category-based Collaborative Filtering algorithm (ECCF). This is is variant of User-to-User Collaborative Filtering that works on category-based user profiles, enriched with preferences inferred from general search behavior.
For the analysis of user interests, we analyzed the query log of a largely used search engine. 

We evaluated ECCF on a large dataset of item ratings, by applying different levels of strictness in the extension of user profiles. The evaluation showed that ECCF outperforms User-to-User Collaborative Filtering in accuracy, MRR, intra-list diversity and user coverage. Interestingly, ECCS also obtains higher accuracy and diversity than the SVD++ recommender system, based on Matrix Factorization; however, ECCS has lower user coverage than SVD++. 

In our future work we will focus on the coverage aspect in order to improve the performance of KNN Collaborative Filtering.
Moreover, we will carry out further experiments, considering (i) a broader domain than Restaurants and Food, on which we have focused our current work, and (ii) users who have provided few or zero ratings. 
We will also analyze other datasets to check whether the performance results described in this article can be generalized. Finally, we will compare the performance of ECCF with a larger set of recommendation approaches based on preference extension.

\begin{acks}
This work was supported by the University of Torino through projects ``Ricerca Locale'', MIMOSA
(MultIModal Ontology-driven query system for the heterogeneous data of a SmArtcity, ``Progetto di Ateneo Torino\_call2014\_L2\_157'', 2015-17)
and the Computer Science PhD program.
We are grateful to Zhongli Filippo Hu, who helped us filter the Yelp dataset.
\end{acks}

  \bibliographystyle{ACM-Reference-Format} 

\begin{thebibliography}{00}


\ifx \showCODEN    \undefined \def \showCODEN     #1{\unskip}     \fi
\ifx \showDOI      \undefined \def \showDOI       #1{{\tt DOI:}\penalty0{#1}\ }
  \fi
\ifx \showISBNx    \undefined \def \showISBNx     #1{\unskip}     \fi
\ifx \showISBNxiii \undefined \def \showISBNxiii  #1{\unskip}     \fi
\ifx \showISSN     \undefined \def \showISSN      #1{\unskip}     \fi
\ifx \showLCCN     \undefined \def \showLCCN      #1{\unskip}     \fi
\ifx \shownote     \undefined \def \shownote      #1{#1}          \fi
\ifx \showarticletitle \undefined \def \showarticletitle #1{#1}   \fi
\ifx \showURL      \undefined \def \showURL       #1{#1}          \fi
\providecommand\bibfield[2]{#2}
\providecommand\bibinfo[2]{#2}
\providecommand\natexlab[1]{#1}
\providecommand\showeprint[2][]{arXiv:#2}

\bibitem[\protect\citeauthoryear{Alotaibi and Vassileva}{Alotaibi and
  Vassileva}{2016}]%
        {Alotaibi-Vassileva:16}
\bibfield{author}{\bibinfo{person}{S. Alotaibi} {and} \bibinfo{person}{J.
  Vassileva}.} \bibinfo{year}{2016}\natexlab{}.
\newblock \showarticletitle{Personalized Recommendation of Research Papers by
  Fusing Recommendations from Explicit and Implicit Social Networks}. In
  \bibinfo{booktitle}{{\em Proc. of {IFUP 2016:} Workshop on Multi-dimensional
  Information Fusion for User Modeling and Personalization}},
  Vol.~\bibinfo{volume}{1618}. \bibinfo{publisher}{{CEUR}},
  \bibinfo{address}{Halifax, Canada}, \bibinfo{pages}{paper 2}.
\newblock


\bibitem[\protect\citeauthoryear{Ardissono, Lucenteforte, Mauro, Savoca,
  Voghera, and {L. La Riccia}}{Ardissono et~al\mbox{.}}{2016}]%
        {Ardissono-etal:16}
\bibfield{author}{\bibinfo{person}{L. Ardissono}, \bibinfo{person}{M.
  Lucenteforte}, \bibinfo{person}{N. Mauro}, \bibinfo{person}{A. Savoca},
  \bibinfo{person}{A. Voghera}, {and} \bibinfo{person}{{L. La Riccia}}.}
  \bibinfo{year}{2016}\natexlab{}.
\newblock \showarticletitle{Exploration of Cultural Heritage Information via
  Textual Search Queries}. In \bibinfo{booktitle}{{\em MobileHCI '16
  Proceedings of the 18th Int. Conf. on Human-Computer Interaction with Mobile
  Devices and Services Adjunct}}. \bibinfo{publisher}{ACM},
  \bibinfo{pages}{992--1001}.
\newblock


\bibitem[\protect\citeauthoryear{Babelfy}{Babelfy}{}]%
        {Babelfy}
\bibfield{author}{\bibinfo{person}{Babelfy}.}
\newblock \showarticletitle{Multilingual Word-sense disambiguation and entity
  linking together!}
\newblock


\bibitem[\protect\citeauthoryear{Bradley and Smyth}{Bradley and Smyth}{2001}]%
        {Bradley-Smyth:01}
\bibfield{author}{\bibinfo{person}{K. Bradley} {and} \bibinfo{person}{B.
  Smyth}.} \bibinfo{year}{2001}\natexlab{}.
\newblock \showarticletitle{Improving Recommendation Diversity}. In
  \bibinfo{booktitle}{{\em {Proc. of the 12th National Conference in Artificial
  Intelligence and Cognitive Science}}},
  \bibfield{editor}{\bibinfo{person}{Diarmuid O'Donoghue}} (Ed.).
  \bibinfo{address}{Maynooth, Ireland}, \bibinfo{pages}{75--84}.
\newblock


\bibitem[\protect\citeauthoryear{Candan, Cataldi, Caro, Sapino, and
  Schifanella}{Candan et~al\mbox{.}}{2009}]%
        {Candan-etal:09}
\bibfield{author}{\bibinfo{person}{{K.S.} Candan}, \bibinfo{person}{M.
  Cataldi}, \bibinfo{person}{L.~Di Caro}, \bibinfo{person}{{M.L.} Sapino},
  {and} \bibinfo{person}{C. Schifanella}.} \bibinfo{year}{2009}\natexlab{}.
\newblock \showarticletitle{{CoSeNa}: a context-based search and navigation
  system}. In \bibinfo{booktitle}{{\em {MEDES '09} Int. Conf. on Management of
  Emergent Digital EcoSystems}}. \bibinfo{address}{Chia Laguna, Italy},
  \bibinfo{pages}{Art. 33}.
\newblock


\bibitem[\protect\citeauthoryear{Cao, Jiang, Pei, He, Liao, Chen, and Li}{Cao
  et~al\mbox{.}}{2008}]%
        {Cao-etal:08}
\bibfield{author}{\bibinfo{person}{H. Cao}, \bibinfo{person}{D. Jiang},
  \bibinfo{person}{J. Pei}, \bibinfo{person}{Q. He}, \bibinfo{person}{Z. Liao},
  \bibinfo{person}{E. Chen}, {and} \bibinfo{person}{H. Li}.}
  \bibinfo{year}{2008}\natexlab{}.
\newblock \showarticletitle{Context-aware query suggestion by mining
  click-through and session data}. In \bibinfo{booktitle}{{\em Proc. of the
  14th {ACM SIGKDD} Int. Conf. on Knowledge Discovery and Data Mining}} {\em
  (\bibinfo{series}{KDD '08})}. \bibinfo{publisher}{ACM}, \bibinfo{address}{New
  York, NY, USA}, \bibinfo{pages}{875--883}.
\newblock


\bibitem[\protect\citeauthoryear{Chen, Chen, and Wang}{Chen
  et~al\mbox{.}}{2015}]%
        {Chen-etal:15}
\bibfield{author}{\bibinfo{person}{L. Chen}, \bibinfo{person}{G. Chen}, {and}
  \bibinfo{person}{F. Wang}.} \bibinfo{year}{2015}\natexlab{}.
\newblock \showarticletitle{Recommender systems based on user reviews: the
  state of the art}.
\newblock \bibinfo{journal}{{\em USER MODELING AND USER-ADAPTED INTERACTION -
  The Journal of Personalization Research\/}} \bibinfo{volume}{25},
  \bibinfo{number}{2} (\bibinfo{year}{2015}), \bibinfo{pages}{99--154}.
\newblock


\bibitem[\protect\citeauthoryear{Chen, Xue, and Yu}{Chen et~al\mbox{.}}{2008}]%
        {Chen-etal:08}
\bibfield{author}{\bibinfo{person}{Y. Chen}, \bibinfo{person}{G-R Xue}, {and}
  \bibinfo{person}{Y. Yu}.} \bibinfo{year}{2008}\natexlab{}.
\newblock \showarticletitle{Advertising Keyword Suggestion Based on Concept
  Hierarchy}. In \bibinfo{booktitle}{{\em Proc. of the 2008 Int. Conf. on Web
  Search and Data Mining}} {\em (\bibinfo{series}{WSDM '08})}.
  \bibinfo{publisher}{ACM}, \bibinfo{address}{New York, NY, USA},
  \bibinfo{pages}{251--260}.
\newblock


\bibitem[\protect\citeauthoryear{Desrosiers and Karypis}{Desrosiers and
  Karypis}{2011}]%
        {Desrosiers-Karypis:11}
\bibfield{author}{\bibinfo{person}{C. Desrosiers} {and} \bibinfo{person}{G.
  Karypis}.} \bibinfo{year}{2011}\natexlab{}.
\newblock \showarticletitle{A Comprehensive Survey of Neighborhood-based
  Recommendation Methods}.
\newblock In \bibinfo{booktitle}{{\em Recommender systems handbook}},
  \bibfield{editor}{\bibinfo{person}{F.~Ricci}, \bibinfo{person}{L.~Rokach},
  \bibinfo{person}{B.~Shapira}, {and} \bibinfo{person}{{P.B.} Kantor}} (Eds.).
  \bibinfo{publisher}{Springer}, \bibinfo{pages}{107--144}.
\newblock


\bibitem[\protect\citeauthoryear{Doychev, Lawlor, Rafter, and Smyth}{Doychev
  et~al\mbox{.}}{2014}]%
        {Doychev-etal:14}
\bibfield{author}{\bibinfo{person}{D. Doychev}, \bibinfo{person}{A. Lawlor},
  \bibinfo{person}{R. Rafter}, {and} \bibinfo{person}{B. Smyth}.}
  \bibinfo{year}{2014}\natexlab{}.
\newblock \showarticletitle{An analysis of recommender algorithms for online
  news}. In \bibinfo{booktitle}{{\em Proc. of {CLEF 2014} Conference and Labs
  of the Evaluation Forum}}. \bibinfo{address}{Sheffield, UK}.
\newblock


\bibitem[\protect\citeauthoryear{Du, Liu, and Jing}{Du et~al\mbox{.}}{2017}]%
        {Du-etal:17}
\bibfield{author}{\bibinfo{person}{Xixi Du}, \bibinfo{person}{Huafeng Liu},
  {and} \bibinfo{person}{Liping Jing}.} \bibinfo{year}{2017}\natexlab{}.
\newblock \showarticletitle{Additive Co-Clustering with Social Influence for
  Recommendation}. In \bibinfo{booktitle}{{\em Proceedings of the Eleventh ACM
  Conference on Recommender Systems}} {\em (\bibinfo{series}{RecSys '17})}.
  \bibinfo{publisher}{ACM}, \bibinfo{address}{New York, NY, USA},
  \bibinfo{pages}{193--200}.
\newblock
\showISBNx{978-1-4503-4652-8}
\showDOI{%
\url{http://dx.doi.org/10.1145/3109859.3109883}}


\bibitem[\protect\citeauthoryear{Fern\'{a}ndez-Tob\'{\i}as, Tomeo, Cantador,
  Noia, and {Di Sciascio}}{Fern\'{a}ndez-Tob\'{\i}as et~al\mbox{.}}{2016}]%
        {Fernandez-Tobias-etal:16}
\bibfield{author}{\bibinfo{person}{I. Fern\'{a}ndez-Tob\'{\i}as},
  \bibinfo{person}{P. Tomeo}, \bibinfo{person}{I. Cantador},
  \bibinfo{person}{T.~Di Noia}, {and} \bibinfo{person}{E. {Di Sciascio}}.}
  \bibinfo{year}{2016}\natexlab{}.
\newblock \showarticletitle{Accuracy and Diversity in Cross-domain
  Recommendations for Cold-start Users with Positive-only Feedback}. In
  \bibinfo{booktitle}{{\em Proc. of the 10th ACM Conference on Recommender
  Systems}} {\em (\bibinfo{series}{RecSys '16})}. \bibinfo{publisher}{ACM},
  \bibinfo{address}{New York, NY, USA}, \bibinfo{pages}{119--122}.
\newblock


\bibitem[\protect\citeauthoryear{Garcin, Dimitrakakis, and Faltings}{Garcin
  et~al\mbox{.}}{2013}]%
        {Garcin-etal:13}
\bibfield{author}{\bibinfo{person}{F. Garcin}, \bibinfo{person}{C.
  Dimitrakakis}, {and} \bibinfo{person}{B Faltings}.}
  \bibinfo{year}{2013}\natexlab{}.
\newblock \showarticletitle{Personalized news recommendation with context
  trees}. In \bibinfo{booktitle}{{\em Proc. of 7th {ACM} Conf. on Recommender
  Systems {(RecSys 2013)}}}. \bibinfo{address}{Honk Kong, China},
  \bibinfo{pages}{105--112}.
\newblock


\bibitem[\protect\citeauthoryear{Gemmel, Schimoler, Mobasher, and Burke}{Gemmel
  et~al\mbox{.}}{2012}]%
        {Gemmel-etal:12}
\bibfield{author}{\bibinfo{person}{J. Gemmel}, \bibinfo{person}{T. Schimoler},
  \bibinfo{person}{B. Mobasher}, {and} \bibinfo{person}{R. Burke}.}
  \bibinfo{year}{2012}\natexlab{}.
\newblock \showarticletitle{Resource recommendation in social annotation
  systems: a linear-weighted hybrid approach}.
\newblock \bibinfo{journal}{{\em Journal of computer and system sciences\/}}
  \bibinfo{volume}{78} (\bibinfo{year}{2012}), \bibinfo{pages}{1160--1174}.
\newblock


\bibitem[\protect\citeauthoryear{Google}{Google}{2017}]%
        {GoogleKnowledgeGraph}
\bibfield{author}{\bibinfo{person}{Google}.} \bibinfo{year}{2017}\natexlab{}.
\newblock \showarticletitle{Knowledge Graph}.
  \bibinfo{address}{\url{https://www.google.com/intl/it\_it/insidesearch/features/search/knowledge.html}}.
\newblock


\bibitem[\protect\citeauthoryear{Greenstein-Messica, Rokach, and
  Friedman}{Greenstein-Messica et~al\mbox{.}}{2017}]%
        {Greenstein-etal:17}
\bibfield{author}{\bibinfo{person}{A. Greenstein-Messica}, \bibinfo{person}{L.
  Rokach}, {and} \bibinfo{person}{M. Friedman}.}
  \bibinfo{year}{2017}\natexlab{}.
\newblock \showarticletitle{Session-based recommendations using item
  embedding}. In \bibinfo{booktitle}{{\em Proc. of the 22nd Int. Conf. on
  Intelligent User Interfaces}} {\em (\bibinfo{series}{IUI '17})}.
  \bibinfo{publisher}{ACM}, \bibinfo{address}{New York, NY, USA},
  \bibinfo{pages}{629--633}.
\newblock


\bibitem[\protect\citeauthoryear{Huang, Chien, and Oyang}{Huang
  et~al\mbox{.}}{2003}]%
        {Huang-etal:03}
\bibfield{author}{\bibinfo{person}{C.-K. Huang}, \bibinfo{person}{L.-F. Chien},
  {and} \bibinfo{person}{Y.-J. Oyang}.} \bibinfo{year}{2003}\natexlab{}.
\newblock \showarticletitle{Relevant term suggestion in interactive web search
  based on contextual in formation in query session logs}.
\newblock \bibinfo{journal}{{\em Journal of the American Society for
  Information Science and Technology\/}} \bibinfo{volume}{54},
  \bibinfo{number}{7} (\bibinfo{year}{2003}), \bibinfo{pages}{638--649}.
\newblock


\bibitem[\protect\citeauthoryear{Hug}{Hug}{2018}]%
        {Surprise}
\bibfield{author}{\bibinfo{person}{Nicolas Hug}.}
  \bibinfo{year}{2018}\natexlab{}.
\newblock \showarticletitle{{S}urprise, a {P}ython library for recommender
  systems}. \bibinfo{howpublished}{\url{http://surpriselib.com}}.
\newblock


\bibitem[\protect\citeauthoryear{Jannach and Ludewig}{Jannach and
  Ludewig}{2017}]%
        {Jannach-Ludewig:17}
\bibfield{author}{\bibinfo{person}{D. Jannach} {and} \bibinfo{person}{M.
  Ludewig}.} \bibinfo{year}{2017}\natexlab{}.
\newblock \showarticletitle{When Recurrent Neural Networks Meet the
  Neighborhood for Session-Based Recommendation}. In \bibinfo{booktitle}{{\em
  Proc. of the Eleventh ACM Conference on Recommender Systems}} {\em
  (\bibinfo{series}{RecSys '17})}. \bibinfo{address}{New York, NY, USA},
  \bibinfo{pages}{306--310}.
\newblock


\bibitem[\protect\citeauthoryear{Jannach, Ludewig, and Lerche}{Jannach
  et~al\mbox{.}}{2017}]%
        {Jannach-etal:17}
\bibfield{author}{\bibinfo{person}{D. Jannach}, \bibinfo{person}{M. Ludewig},
  {and} \bibinfo{person}{L. Lerche}.} \bibinfo{year}{2017}\natexlab{}.
\newblock \showarticletitle{Session-based Item Recommendation in e-Commerce: On
  Short-term Intents, Reminders, Trends and Discounts}.
\newblock \bibinfo{journal}{{\em User Modeling and User-Adapted Interaction\/}}
  \bibinfo{volume}{27}, \bibinfo{number}{3-5} (\bibinfo{year}{2017}),
  \bibinfo{pages}{351--392}.
\newblock


\bibitem[\protect\citeauthoryear{Koren}{Koren}{2008}]%
        {Koren:08}
\bibfield{author}{\bibinfo{person}{Yehuda Koren}.}
  \bibinfo{year}{2008}\natexlab{}.
\newblock \showarticletitle{Factorization Meets the Neighborhood: A
  Multifaceted Collaborative Filtering Model}. In \bibinfo{booktitle}{{\em
  Proceedings of the 14th ACM SIGKDD International Conference on Knowledge
  Discovery and Data Mining}} {\em (\bibinfo{series}{KDD '08})}.
  \bibinfo{publisher}{ACM}, \bibinfo{address}{New York, NY, USA},
  \bibinfo{pages}{426--434}.
\newblock
\showISBNx{978-1-60558-193-4}
\showDOI{%
\url{http://dx.doi.org/10.1145/1401890.1401944}}


\bibitem[\protect\citeauthoryear{Koren and Bell}{Koren and Bell}{2011}]%
        {Koren-Bell:11}
\bibfield{author}{\bibinfo{person}{Y. Koren} {and} \bibinfo{person}{R. Bell}.}
  \bibinfo{year}{2011}\natexlab{}.
\newblock \showarticletitle{Advances in collaborative filtering}.
\newblock In \bibinfo{booktitle}{{\em Recommender systems handbook}},
  \bibfield{editor}{\bibinfo{person}{F.~Ricci}, \bibinfo{person}{L.~Rokach},
  \bibinfo{person}{B.~Shapira}, {and} \bibinfo{person}{{P.B.} Kantor}} (Eds.).
  \bibinfo{publisher}{Springer}, \bibinfo{pages}{145--186}.
\newblock


\bibitem[\protect\citeauthoryear{Kuter and Golbeck}{Kuter and Golbeck}{2007}]%
        {Kuter-etal:07}
\bibfield{author}{\bibinfo{person}{U. Kuter} {and} \bibinfo{person}{J.
  Golbeck}.} \bibinfo{year}{2007}\natexlab{}.
\newblock \showarticletitle{SUNNY: A New Algorithm for Trust Inference in
  Social Networks Using Probabilistic Confidence Models}. In
  \bibinfo{booktitle}{{\em Proc. of the 22Nd National Conference on Artificial
  Intelligence - Volume 2}} {\em (\bibinfo{series}{AAAI'07})}.
  \bibinfo{publisher}{AAAI Press}, \bibinfo{pages}{1377--1382}.
\newblock


\bibitem[\protect\citeauthoryear{Liu and Lee}{Liu and Lee}{2010}]%
        {Liu-Lee:10}
\bibfield{author}{\bibinfo{person}{Fengkun Liu} {and} \bibinfo{person}{Hong~Joo
  Lee}.} \bibinfo{year}{2010}\natexlab{}.
\newblock \showarticletitle{Use of Social Network Information to Enhance
  Collaborative Filtering Performance}.
\newblock \bibinfo{journal}{{\em Expert Syst. Appl.\/}} \bibinfo{volume}{37},
  \bibinfo{number}{7} (\bibinfo{date}{July} \bibinfo{year}{2010}),
  \bibinfo{pages}{4772--4778}.
\newblock
\showISSN{0957-4174}
\showDOI{%
\url{http://dx.doi.org/10.1016/j.eswa.2009.12.061}}


\bibitem[\protect\citeauthoryear{Lu, Dong, and Smyth}{Lu et~al\mbox{.}}{2018}]%
        {Lu-etal:18}
\bibfield{author}{\bibinfo{person}{Y. Lu}, \bibinfo{person}{R. Dong}, {and}
  \bibinfo{person}{B. Smyth}.} \bibinfo{year}{2018}\natexlab{}.
\newblock \showarticletitle{Coevolutionary Recommendation Model: Mutual
  Learning Between Ratings and Reviews}. In \bibinfo{booktitle}{{\em Proc. of
  the 2018 World Wide Web Conference}} {\em (\bibinfo{series}{WWW '18})}.
  \bibinfo{publisher}{International World Wide Web Conferences Steering
  Committee}, \bibinfo{address}{Republic and Canton of Geneva, Switzerland},
  \bibinfo{pages}{773--782}.
\newblock


\bibitem[\protect\citeauthoryear{Ludewig and Jannach}{Ludewig and
  Jannach}{2018}]%
        {Ludewig-Jannach:18}
\bibfield{author}{\bibinfo{person}{M. Ludewig} {and} \bibinfo{person}{D.
  Jannach}.} \bibinfo{year}{2018}\natexlab{}.
\newblock \showarticletitle{Evaluation of Session-based Recommendation
  Algorithms}.
\newblock \bibinfo{journal}{{\em User-Modeling and User-Adapted Interaction\/}}
  (\bibinfo{year}{2018}).
\newblock


\bibitem[\protect\citeauthoryear{Mauro and Ardissono}{Mauro and
  Ardissono}{2017}]%
        {Mauro-Ardissono:17b}
\bibfield{author}{\bibinfo{person}{N. Mauro} {and} \bibinfo{person}{L.
  Ardissono}.} \bibinfo{year}{2017}\natexlab{}.
\newblock \showarticletitle{Concept-aware Geographic Information Retrieval}. In
  \bibinfo{booktitle}{{\em Proc. of 2017 IEEE/WIC/ACM Int. Conf. on Web
  Intelligence (WI)}}. \bibinfo{publisher}{ACM}, \bibinfo{address}{Leipzig,
  Germany}, \bibinfo{pages}{34--41}.
\newblock


\bibitem[\protect\citeauthoryear{Mauro and Ardissono}{Mauro and
  Ardissono}{2018}]%
        {Mauro-Ardissono:18}
\bibfield{author}{\bibinfo{person}{N. Mauro} {and} \bibinfo{person}{L.
  Ardissono}.} \bibinfo{year}{2018}\natexlab{}.
\newblock \showarticletitle{Session-based Suggestion of Topics for Exploratory
  Search}. In \bibinfo{booktitle}{{\em Proc. of {ACM IUI 2018}}}.
  \bibinfo{publisher}{ACM}, \bibinfo{address}{Tokyo, Japan},
  \bibinfo{pages}{341--352}.
\newblock


\bibitem[\protect\citeauthoryear{Mauro, Ardissono, and Hu}{Mauro
  et~al\mbox{.}}{2019}]%
        {Mauro-etal:19}
\bibfield{author}{\bibinfo{person}{N. Mauro}, \bibinfo{person}{L. Ardissono},
  {and} \bibinfo{person}{{Z.F.} Hu}.} \bibinfo{year}{2019}\natexlab{}.
\newblock \showarticletitle{Multi-faceted Trust-based Collaborative Filtering}.
  In \bibinfo{booktitle}{{\em Proc. of {ACM UMAP 2019}}}.
  \bibinfo{publisher}{ACM}, \bibinfo{address}{Larnaca, Cyprus},
  \bibinfo{pages}{to appear}.
\newblock


\bibitem[\protect\citeauthoryear{Mcnally, O'Mahony, and Smyth}{Mcnally
  et~al\mbox{.}}{2014}]%
        {Mcnally-etal:14}
\bibfield{author}{\bibinfo{person}{Kevin Mcnally}, \bibinfo{person}{Michael~P.
  O'Mahony}, {and} \bibinfo{person}{Barry Smyth}.}
  \bibinfo{year}{2014}\natexlab{}.
\newblock \showarticletitle{A Comparative Study of Collaboration-based
  Reputation Models for Social Recommender Systems}.
\newblock \bibinfo{journal}{{\em User Modeling and User-Adapted Interaction\/}}
  \bibinfo{volume}{24}, \bibinfo{number}{3} (\bibinfo{date}{Aug.}
  \bibinfo{year}{2014}), \bibinfo{pages}{219--260}.
\newblock
\showISSN{0924-1868}
\showDOI{%
\url{http://dx.doi.org/10.1007/s11257-013-9143-6}}


\bibitem[\protect\citeauthoryear{Muhammad, Lawlor, Rafter, and Smyth}{Muhammad
  et~al\mbox{.}}{2015}]%
        {Muhammad-etal:15}
\bibfield{author}{\bibinfo{person}{K. Muhammad}, \bibinfo{person}{A. Lawlor},
  \bibinfo{person}{E. Rafter}, {and} \bibinfo{person}{B. Smyth}.}
  \bibinfo{year}{2015}\natexlab{}.
\newblock \showarticletitle{Great Explanations: Opinionated Explanations for
  Recommendations}. In \bibinfo{booktitle}{{\em Case-Based Reasoning Research
  and Development}}, \bibfield{editor}{\bibinfo{person}{Eyke H{\"u}llermeier}
  {and} \bibinfo{person}{Mirjam Minor}} (Eds.). \bibinfo{publisher}{Springer
  International Publishing}, \bibinfo{address}{Cham},
  \bibinfo{pages}{244--258}.
\newblock


\bibitem[\protect\citeauthoryear{Musat and Faltings}{Musat and
  Faltings}{2015}]%
        {Musat-Faltings:15}
\bibfield{author}{\bibinfo{person}{{C.-C.} Musat} {and} \bibinfo{person}{B.
  Faltings}.} \bibinfo{year}{2015}\natexlab{}.
\newblock \showarticletitle{Personalizing product rankings using collaborative
  filtering on opinion-derived topic profiles}. In \bibinfo{booktitle}{{\em
  Proc. 24th {IJCAI}}}. \bibinfo{address}{Buenos Aires, Argentina},
  \bibinfo{pages}{830--836}.
\newblock


\bibitem[\protect\citeauthoryear{Musto, Basile, Lops, {de Gemmis}, and
  Semeraro}{Musto et~al\mbox{.}}{2017}]%
        {Musto-etal:16}
\bibfield{author}{\bibinfo{person}{C. Musto}, \bibinfo{person}{P. Basile},
  \bibinfo{person}{P. Lops}, \bibinfo{person}{M. {de Gemmis}}, {and}
  \bibinfo{person}{G. Semeraro}.} \bibinfo{year}{2017}\natexlab{}.
\newblock \showarticletitle{Introducing Linked Open Data in Graph-based
  Recommender Systems}.
\newblock \bibinfo{journal}{{\em Inf. Process. Manage.\/}}
  \bibinfo{volume}{53}, \bibinfo{number}{2} (\bibinfo{year}{2017}),
  \bibinfo{pages}{405--435}.
\newblock


\bibitem[\protect\citeauthoryear{Musto, Franza, Semeraro, {de Gemmis}, and
  Lops}{Musto et~al\mbox{.}}{2018}]%
        {Musto-etal:18}
\bibfield{author}{\bibinfo{person}{C. Musto}, \bibinfo{person}{T. Franza},
  \bibinfo{person}{G. Semeraro}, \bibinfo{person}{M. {de Gemmis}}, {and}
  \bibinfo{person}{P. Lops}.} \bibinfo{year}{2018}\natexlab{}.
\newblock \showarticletitle{Deep Content-based Recommender Systems Exploiting
  Recurrent Neural Networks and Linked Open Data}. In \bibinfo{booktitle}{{\em
  Adjunct Publication of the 26th Conference on User Modeling, Adaptation and
  Personalization}} {\em (\bibinfo{series}{UMAP '18})}.
  \bibinfo{publisher}{ACM}, \bibinfo{address}{New York, NY, USA},
  \bibinfo{pages}{239--244}.
\newblock


\bibitem[\protect\citeauthoryear{Musto, Lops, {de Gemmis}, and Semeraro}{Musto
  et~al\mbox{.}}{2017}]%
        {Musto-etal:17}
\bibfield{author}{\bibinfo{person}{C. Musto}, \bibinfo{person}{P. Lops},
  \bibinfo{person}{M. {de Gemmis}}, {and} \bibinfo{person}{G. Semeraro}.}
  \bibinfo{year}{2017}\natexlab{}.
\newblock \showarticletitle{Semantics-aware Recommender Systems exploiting
  Linked Open Data and graph-based features}.
\newblock \bibinfo{journal}{{\em Knowledge-Based Systems\/}}
  \bibinfo{volume}{136} (\bibinfo{year}{2017}), \bibinfo{pages}{1 -- 14}.
\newblock


\bibitem[\protect\citeauthoryear{Musto, Semeraro, Lovascio, de~Gemmis, and
  Lops}{Musto et~al\mbox{.}}{2018}]%
        {Musto-etal:2018b}
\bibfield{author}{\bibinfo{person}{Cataldo Musto}, \bibinfo{person}{Giovanni
  Semeraro}, \bibinfo{person}{Cosimo Lovascio}, \bibinfo{person}{Marco de
  Gemmis}, {and} \bibinfo{person}{Pasquale Lops}.}
  \bibinfo{year}{2018}\natexlab{}.
\newblock \showarticletitle{A Framework for Holistic User Modeling Merging
  Heterogeneous Digital Footprints}. In \bibinfo{booktitle}{{\em Adj. Publ. of
  {UMAP '18}}}. \bibinfo{publisher}{ACM}, \bibinfo{address}{New York, USA},
  \bibinfo{pages}{97--101}.
\newblock


\bibitem[\protect\citeauthoryear{Nakamoto, Nakajima, Miyazaki, and
  Uemura}{Nakamoto et~al\mbox{.}}{2007}]%
        {Nakamoto:2007}
\bibfield{author}{\bibinfo{person}{Reyn Nakamoto}, \bibinfo{person}{Shinsuke
  Nakajima}, \bibinfo{person}{Jun Miyazaki}, {and} \bibinfo{person}{Shunsuke
  Uemura}.} \bibinfo{year}{2007}\natexlab{}.
\newblock \showarticletitle{Tag-based contextual collaborative filtering.}
\newblock \bibinfo{journal}{{\em IAENG Int. Journal of Computer Science\/}}
  \bibinfo{volume}{34}, \bibinfo{number}{2} (\bibinfo{year}{2007}).
\newblock


\bibitem[\protect\citeauthoryear{Noia, Ostuni, Rosati, Tomeo, {Di Sciascio},
  Mirizzi, and Bartolini}{Noia et~al\mbox{.}}{2016}]%
        {DiNoia-etal:16}
\bibfield{author}{\bibinfo{person}{T.~Di Noia}, \bibinfo{person}{{V.C.}
  Ostuni}, \bibinfo{person}{J. Rosati}, \bibinfo{person}{P. Tomeo},
  \bibinfo{person}{E. {Di Sciascio}}, \bibinfo{person}{R. Mirizzi}, {and}
  \bibinfo{person}{C. Bartolini}.} \bibinfo{year}{2016}\natexlab{}.
\newblock \showarticletitle{Building a Relatedness Graph from Linked Open
  Data}.
\newblock \bibinfo{journal}{{\em Expert Syst. Appl.\/}} \bibinfo{volume}{44},
  \bibinfo{number}{C} (\bibinfo{year}{2016}), \bibinfo{pages}{354--366}.
\newblock


\bibitem[\protect\citeauthoryear{Oramas, Ostuni, {Di Noia}, Serra, and {Di
  Sciascio}}{Oramas et~al\mbox{.}}{2015}]%
        {Oramas-etal:15}
\bibfield{author}{\bibinfo{person}{S. Oramas}, \bibinfo{person}{{V.C.} Ostuni},
  \bibinfo{person}{T. {Di Noia}}, \bibinfo{person}{X. Serra}, {and}
  \bibinfo{person}{E. {Di Sciascio}}.} \bibinfo{year}{2015}\natexlab{}.
\newblock \showarticletitle{Sound and music recommendation with knowledge
  graphs}.
\newblock \bibinfo{journal}{{\em {ACM} Transactions on Intelligent Systems and
  Technology\/}} \bibinfo{volume}{8}, \bibinfo{number}{2}
  (\bibinfo{year}{2015}), \bibinfo{pages}{Art. 21}.
\newblock


\bibitem[\protect\citeauthoryear{Palumbo, Rizzo, and Troncy}{Palumbo
  et~al\mbox{.}}{2017}]%
        {Palumbo-etal:17}
\bibfield{author}{\bibinfo{person}{E. Palumbo}, \bibinfo{person}{G. Rizzo},
  {and} \bibinfo{person}{R. Troncy}.} \bibinfo{year}{2017}\natexlab{}.
\newblock \showarticletitle{Entity2Rec: Learning User-Item Relatedness from
  Knowledge Graphs for Top-N Item Recommendation}. In \bibinfo{booktitle}{{\em
  Proc. of the Eleventh ACM Conference on Recommender Systems}} {\em
  (\bibinfo{series}{RecSys '17})}. \bibinfo{publisher}{ACM},
  \bibinfo{address}{New York, NY, USA}, \bibinfo{pages}{32--36}.
\newblock


\bibitem[\protect\citeauthoryear{Ragone, Tomeo, Magarelli, {Di Noia},
  Palmonari, Maurino, and {Di Sciascio}}{Ragone et~al\mbox{.}}{2017}]%
        {Ragone-etal:17}
\bibfield{author}{\bibinfo{person}{A. Ragone}, \bibinfo{person}{P. Tomeo},
  \bibinfo{person}{C. Magarelli}, \bibinfo{person}{T. {Di Noia}},
  \bibinfo{person}{M. Palmonari}, \bibinfo{person}{A. Maurino}, {and}
  \bibinfo{person}{E. {Di Sciascio}}.} \bibinfo{year}{2017}\natexlab{}.
\newblock \showarticletitle{Schema-summarization in Linked-data-based Feature
  Selection for Recommender Systems}. In \bibinfo{booktitle}{{\em Proc. of the
  Symposium on Applied Computing}} {\em (\bibinfo{series}{SAC '17})}.
  \bibinfo{publisher}{ACM}, \bibinfo{address}{New York, NY, USA},
  \bibinfo{pages}{330--335}.
\newblock


\bibitem[\protect\citeauthoryear{Ricci, Rokach, and B.~Shapira}{Ricci
  et~al\mbox{.}}{2011}]%
        {Ricci-etal:11}
\bibfield{author}{\bibinfo{person}{F. Ricci}, \bibinfo{person}{L. Rokach},
  {and} \bibinfo{person}{Bracha B.~Shapira}.} \bibinfo{year}{2011}\natexlab{}.
\newblock \bibinfo{booktitle}{{\em Introduction to Recommender Systems
  Handbook}}.
\newblock \bibinfo{publisher}{Springer US}, \bibinfo{address}{Boston, MA},
  \bibinfo{pages}{1--35}.
\newblock


\bibitem[\protect\citeauthoryear{Rieh and Xie}{Rieh and Xie}{2006}]%
        {Rieh-Xie:06}
\bibfield{author}{\bibinfo{person}{{S.Y.} Rieh} {and} \bibinfo{person}{{H. I}
  Xie}.} \bibinfo{year}{2006}\natexlab{}.
\newblock \showarticletitle{Analysis of multiple query reformulations on the
  web: the interactive information retrieval context}.
\newblock \bibinfo{journal}{{\em Information processing and management\/}}
  \bibinfo{volume}{42} (\bibinfo{year}{2006}), \bibinfo{pages}{751--768}.
\newblock


\bibitem[\protect\citeauthoryear{Ronen, {Yom-Tov}, and Lavee}{Ronen
  et~al\mbox{.}}{2016}]%
        {Ronen-etal:16}
\bibfield{author}{\bibinfo{person}{R. Ronen}, \bibinfo{person}{E. {Yom-Tov}},
  {and} \bibinfo{person}{G. Lavee}.} \bibinfo{year}{2016}\natexlab{}.
\newblock \showarticletitle{Recommendations meet web browsing: enhancing
  collaborative filtering using internet browsing logs}. In
  \bibinfo{booktitle}{{\em Proc.of 2016 {IEEE} 32nd International Conference on
  Data Engineering ({ICDE})}}. \bibinfo{address}{Kelsinki, Finland},
  \bibinfo{pages}{1230--1238}.
\newblock


\bibitem[\protect\citeauthoryear{Sen, Vig, and Riedl}{Sen
  et~al\mbox{.}}{2009}]%
        {Sen-etal:09}
\bibfield{author}{\bibinfo{person}{S. Sen}, \bibinfo{person}{J. Vig}, {and}
  \bibinfo{person}{J. Riedl}.} \bibinfo{year}{2009}\natexlab{}.
\newblock \showarticletitle{Tagommenders: Connecting Users to Items Through
  Tags}. In \bibinfo{booktitle}{{\em Proc. of the 18th International Conference
  on World Wide Web}} {\em (\bibinfo{series}{WWW '09})}.
  \bibinfo{publisher}{ACM}, \bibinfo{address}{New York, NY, USA},
  \bibinfo{pages}{671--680}.
\newblock


\bibitem[\protect\citeauthoryear{Sieg, Mobasher, and Burke}{Sieg
  et~al\mbox{.}}{2007}]%
        {Sieg-etal:07b}
\bibfield{author}{\bibinfo{person}{A. Sieg}, \bibinfo{person}{B. Mobasher},
  {and} \bibinfo{person}{R. Burke}.} \bibinfo{year}{2007}\natexlab{}.
\newblock \showarticletitle{Web Search Personalization with Ontological User
  Profiles}. In \bibinfo{booktitle}{{\em Proc. of the 16th ACM Conf. on
  Information and Knowledge Management}} {\em (\bibinfo{series}{CIKM '07})}.
  \bibinfo{publisher}{ACM}, \bibinfo{address}{New York, NY, USA},
  \bibinfo{pages}{525--534}.
\newblock


\bibitem[\protect\citeauthoryear{Sieg, Mobasher, and Burke}{Sieg
  et~al\mbox{.}}{2010}]%
        {Sieg-etal:10b}
\bibfield{author}{\bibinfo{person}{A. Sieg}, \bibinfo{person}{B. Mobasher},
  {and} \bibinfo{person}{R. Burke}.} \bibinfo{year}{2010}\natexlab{}.
\newblock \showarticletitle{Ontology-based collaborative recommendation}.
\newblock \bibinfo{journal}{{\em Computing\/}} (\bibinfo{year}{2010}).
\newblock


\bibitem[\protect\citeauthoryear{Tang, Hu, Gao, and Liu}{Tang
  et~al\mbox{.}}{2013}]%
        {Tang-etal:13}
\bibfield{author}{\bibinfo{person}{Jiliang Tang}, \bibinfo{person}{Xia Hu},
  \bibinfo{person}{Huiji Gao}, {and} \bibinfo{person}{Huan Liu}.}
  \bibinfo{year}{2013}\natexlab{}.
\newblock \showarticletitle{Exploiting Local and Global Social Context for
  Recommendation}. In \bibinfo{booktitle}{{\em Proceedings of the Twenty-Third
  International Joint Conference on Artificial Intelligence}} {\em
  (\bibinfo{series}{IJCAI '13})}. \bibinfo{publisher}{AAAI Press},
  \bibinfo{pages}{2712--2718}.
\newblock
\showISBNx{978-1-57735-633-2}
\showURL{%
\url{http://dl.acm.org/citation.cfm?id=2540128.2540519}}


\bibitem[\protect\citeauthoryear{Teevan, Dumais, and Horvitz}{Teevan
  et~al\mbox{.}}{2005}]%
        {Teevan-etal:05}
\bibfield{author}{\bibinfo{person}{J. Teevan}, \bibinfo{person}{{S.T.} Dumais},
  {and} \bibinfo{person}{E. Horvitz}.} \bibinfo{year}{2005}\natexlab{}.
\newblock \showarticletitle{Personalizing Search via Automated Analysis of
  Interests and Activities}. In \bibinfo{booktitle}{{\em Proc. of the 28th
  Annual Int. ACM SIGIR Conf. on Research and Development in Information
  Retrieval}} {\em (\bibinfo{series}{SIGIR '05})}. \bibinfo{publisher}{ACM},
  \bibinfo{address}{New York, NY, USA}, \bibinfo{pages}{449--456}.
\newblock


\bibitem[\protect\citeauthoryear{Tso-Sutter, Marinho, and
  Schmidt-Thieme}{Tso-Sutter et~al\mbox{.}}{2008}]%
        {Tso-Sutter:2008}
\bibfield{author}{\bibinfo{person}{Karen H.~L. Tso-Sutter},
  \bibinfo{person}{Leandro~Balby Marinho}, {and} \bibinfo{person}{Lars
  Schmidt-Thieme}.} \bibinfo{year}{2008}\natexlab{}.
\newblock \showarticletitle{Tag-aware Recommender Systems by Fusion of
  Collaborative Filtering Algorithms}. In \bibinfo{booktitle}{{\em Proc. of
  {SAC '08}}}. \bibinfo{publisher}{ACM}, \bibinfo{address}{New York, NY, USA},
  \bibinfo{pages}{1995--1999}.
\newblock


\bibitem[\protect\citeauthoryear{Vahedian, Burke, and Mobasher}{Vahedian
  et~al\mbox{.}}{2017}]%
        {Vahedian-etal:17}
\bibfield{author}{\bibinfo{person}{Fatemeh Vahedian}, \bibinfo{person}{Robin
  Burke}, {and} \bibinfo{person}{Bamshad Mobasher}.}
  \bibinfo{year}{2017}\natexlab{}.
\newblock \showarticletitle{Multirelational Recommendation in Heterogeneous
  Networks}.
\newblock \bibinfo{journal}{{\em ACM Trans. Web\/}} \bibinfo{volume}{11},
  \bibinfo{number}{3}, Article \bibinfo{articleno}{15} (\bibinfo{year}{2017}),
  \bibinfo{numpages}{34}~pages.
\newblock


\bibitem[\protect\citeauthoryear{White, Bilenko, and Cucerzan}{White
  et~al\mbox{.}}{2007}]%
        {White-etal:07}
\bibfield{author}{\bibinfo{person}{{R.W.} White}, \bibinfo{person}{M. Bilenko},
  {and} \bibinfo{person}{S. Cucerzan}.} \bibinfo{year}{2007}\natexlab{}.
\newblock \showarticletitle{Studying the use of popular destinations to enhance
  web search interaction}. In \bibinfo{booktitle}{{\em Proc. of the 30th Annual
  Int. ACM SIGIR Conf. on Research and Development in Information Retrieval}}
  {\em (\bibinfo{series}{SIGIR '07})}. \bibinfo{publisher}{ACM},
  \bibinfo{address}{New York, NY, USA}, \bibinfo{pages}{159--166}.
\newblock


\bibitem[\protect\citeauthoryear{{WordNet}}{{WordNet}}{2017}]%
        {WordNet}
\bibfield{author}{\bibinfo{person}{{WordNet}}.}
  \bibinfo{year}{2017}\natexlab{}.
\newblock \showarticletitle{{WordNet} - a lexical database for {English}}.
  \bibinfo{address}{\url{https://wordnet.princeton.edu/}}.
\newblock


\bibitem[\protect\citeauthoryear{Wu, Li, Wang, and Zhu}{Wu
  et~al\mbox{.}}{2012}]%
        {Wu-etal:12}
\bibfield{author}{\bibinfo{person}{W. Wu}, \bibinfo{person}{H. Li},
  \bibinfo{person}{H. Wang}, {and} \bibinfo{person}{{K.Q} Zhu}.}
  \bibinfo{year}{2012}\natexlab{}.
\newblock \showarticletitle{Probase: A Probabilistic Taxonomy for Text
  Understanding}. In \bibinfo{booktitle}{{\em Proc. of the 2012 ACM SIGMOD Int.
  Conf. on Management of Data}}. \bibinfo{publisher}{ACM},
  \bibinfo{address}{New York, NY, USA}, \bibinfo{pages}{481--492}.
\newblock


\bibitem[\protect\citeauthoryear{Yang, Lei, Liu, and Li}{Yang
  et~al\mbox{.}}{2017}]%
        {Yang-etal:17}
\bibfield{author}{\bibinfo{person}{B. Yang}, \bibinfo{person}{Y. Lei},
  \bibinfo{person}{J. Liu}, {and} \bibinfo{person}{W. Li}.}
  \bibinfo{year}{2017}\natexlab{}.
\newblock \showarticletitle{Social collaborative filtering by trust}.
\newblock \bibinfo{journal}{{\em {IEEE} Transactions on Pattern Analysis and
  Machine Intelligence\/}} \bibinfo{volume}{39}, \bibinfo{number}{8}
  (\bibinfo{year}{2017}), \bibinfo{pages}{1633--1647}.
\newblock


\bibitem[\protect\citeauthoryear{Yelp}{Yelp}{}]%
        {Yelp-dataset}
\bibfield{author}{\bibinfo{person}{Yelp}.}
\newblock \showarticletitle{Yelp Dataset Challenge}.
  \bibinfo{address}{\url{https://www.yelp.com/dataset\_challenge}}.
\newblock


\bibitem[\protect\citeauthoryear{Zhen, Li, and Yeung}{Zhen
  et~al\mbox{.}}{2009}]%
        {Zhen:2009}
\bibfield{author}{\bibinfo{person}{Yi Zhen}, \bibinfo{person}{Wu-Jun Li}, {and}
  \bibinfo{person}{Dit-Yan Yeung}.} \bibinfo{year}{2009}\natexlab{}.
\newblock \showarticletitle{TagiCoFi: Tag Informed Collaborative Filtering}. In
  \bibinfo{booktitle}{{\em Proc. of {RecSys '09}}}. \bibinfo{publisher}{ACM},
  \bibinfo{pages}{69--76}.
\newblock


\end{thebibliography}
\balance

\end{document}